\documentclass[prl,twocolumn,showpacs,superscriptaddress,floatfix,reprint]{revtex4-1}

\usepackage[utf8]{inputenc}
\usepackage[english]{babel}
\usepackage[T1]{fontenc}
\usepackage{graphicx}
\usepackage{amsmath}
\usepackage{amsfonts}
\usepackage{amssymb}
\usepackage{siunitx}
\usepackage{textcomp}
\usepackage{standalone}
\usepackage{xspace}
\usepackage[caption=false]{subfig}
\usepackage[normalem]{ulem}
\usepackage{color}

\newcommand{\new}[1]{\textcolor{black}{#1}}

\begin{document}

\title{Optimizing Hyperuniformity in Self-assembled Bidisperse Emulsions}
\author{Joshua Ricouvier}
\affiliation{ESPCI Paris, PSL Research University, CNRS, IPGG, MMN, 6 rue Jean Calvin, F-75005, Paris, France}
\author{Romain Pierrat}
\affiliation{ESPCI Paris, PSL Research University, CNRS, Institut Langevin, 1 rue Jussieu, F-75005, Paris, France}
\author{Rémi Carminati}
\affiliation{ESPCI Paris, PSL Research University, CNRS, Institut Langevin, 1 rue Jussieu, F-75005, Paris, France}
\author{Patrick Tabeling}
\affiliation{ESPCI Paris, PSL Research University, CNRS, IPGG, MMN, 6 rue Jean Calvin, F-75005, Paris, France}
\author{Pavel Yazhgur}
\affiliation{ESPCI Paris, PSL Research University, CNRS, IPGG, MMN, 6 rue Jean Calvin, F-75005, Paris, France}

\begin{abstract}
   We study long range density fluctuations (hyperuniformity) in two-dimensional jammed packings of bidisperse droplets.
   Taking advantage of microfuidics, we systematically span a large range of size and concentration ratios of the two
   droplet populations. We identify various defects increasing long range density fluctuations mainly due to
   organization of local particle environment. By choosing an appropriate bidispersity, we fabricate materials with a
   high level of hyperuniformity. Interesting transparency properties of these optimized materials are established based
   on numerical simulations.
\end{abstract}

\pacs{47.57.Bc, 05.65.+b, 82.70.Kj}

\maketitle

Hyperuniform systems have been recently identified as a particular subclass of disordered systems with \new{density
fluctuations} vanishing at infinitely large length scales~\cite{torquato2003local}. In Fourier space, the structure
factor $S(q)$ reveals hyperuniformity by vanishing when $q$ tends to zero. These systems possess  properties of
crystals, such as long-range order, in addition to characteristics of disordered systems such as a statistically
isotropic structure. Hyperuniformity has been used as a guideline to create disordered photonic band gap materials
(PBG)~\cite{florescu2009designer,muller2014silicon,muller2017photonic}. It paves the way for fabricating colloidal
materials with forbidden band gaps less sensitive to defects than those observed in ordered colloidal arrangements.
Disorder also brings isotropy, useful to create free-form waveguides, laser cavities, noniridescent dyes or display
devices~\cite{man2013isotropic,park2014full,degl2016hyperuniform}. It was also shown that hyperuniform materials can be
optimized to be optically dense and transparent at the same time~\cite{Leseur:16}. The obtained results have been naturally 
extended to other types of propagating waves, for example to produce disordered phononic or
electronic band gaps \cite{gkantzounis2017hyperuniform,hejna2013nearly}. Such active hyperuniform materials are now
mainly designed on a computer and are printed 
afterwards~\cite{man2013photonic,muller2014silicon,muller2017photonic,SCHEFFOLD-2013}.  These patterns can be optimized
to exhibit tremendous complete photonic band gap but suffer from severe limitations of 3D printing (time scale, defects,
resolution, etc). By contrast, the bottom-up approach, potentially more interesting in terms of throughput and cost, has
not yet led to innovative materials~\cite{weijs2015emergent,dreyfus2015diagnosing}.

Using numerical simulations, Donev \emph{et al.} predicted~\cite{PhysRevLett.95.090604} that random close packing of
spheres can be hyperuniform under certain conditions. Such obtained structures are called Maximally Random Jammed (MRJ)
states and are defined as the most disordered strictly jammed packing of solid spheres~\cite{torquato2000random} which
minimizes an arbitrary order parameter, e.g. a fraction of crystallized particles. To realize such assemblies, the
system should avoid two kinds of defects: rattlers (particles free to move in a confining cage) and crystalline domains.
The former refers to zones less dense than the surrounding medium (underpacked defects), while the latter are denser
(overpacked defects)~\cite{rieser2016divergence}. Both create long range density fluctuations and are unfavorable for
hyperuniformity.

In this Letter, we demonstrate the creation of hyperuniform structures by self-assembling bidisperse droplets in a
Hele-Shaw configuration using a microfluidic chip. The small friction between liquid droplets, together with
deformability, allow us to close the cages around rattling particles and create truly jammed states, while size 
bidispersity breaks the crystallisation. We take advantage of microfluidics to produce a
set of particle ensembles in typically a few minutes,  precisely controlling the size and number ratios of droplets. We
thus systematically vary both ratios, unlike most studies of jammed states. Interestingly, we discover that
hyperuniformity defects can be much more complex than expected. Density fluctuations turn out to be very sensitive to
the structure of clusters formed by the particles and their immediate neighbors. Thus a whole new way to describe
hyperuniform jammed systems is proposed. We believe that this can guide \new{research} for new materials with vanishing
density fluctuations. 

\begin{figure}[t]
   \includegraphics[width=1\linewidth]{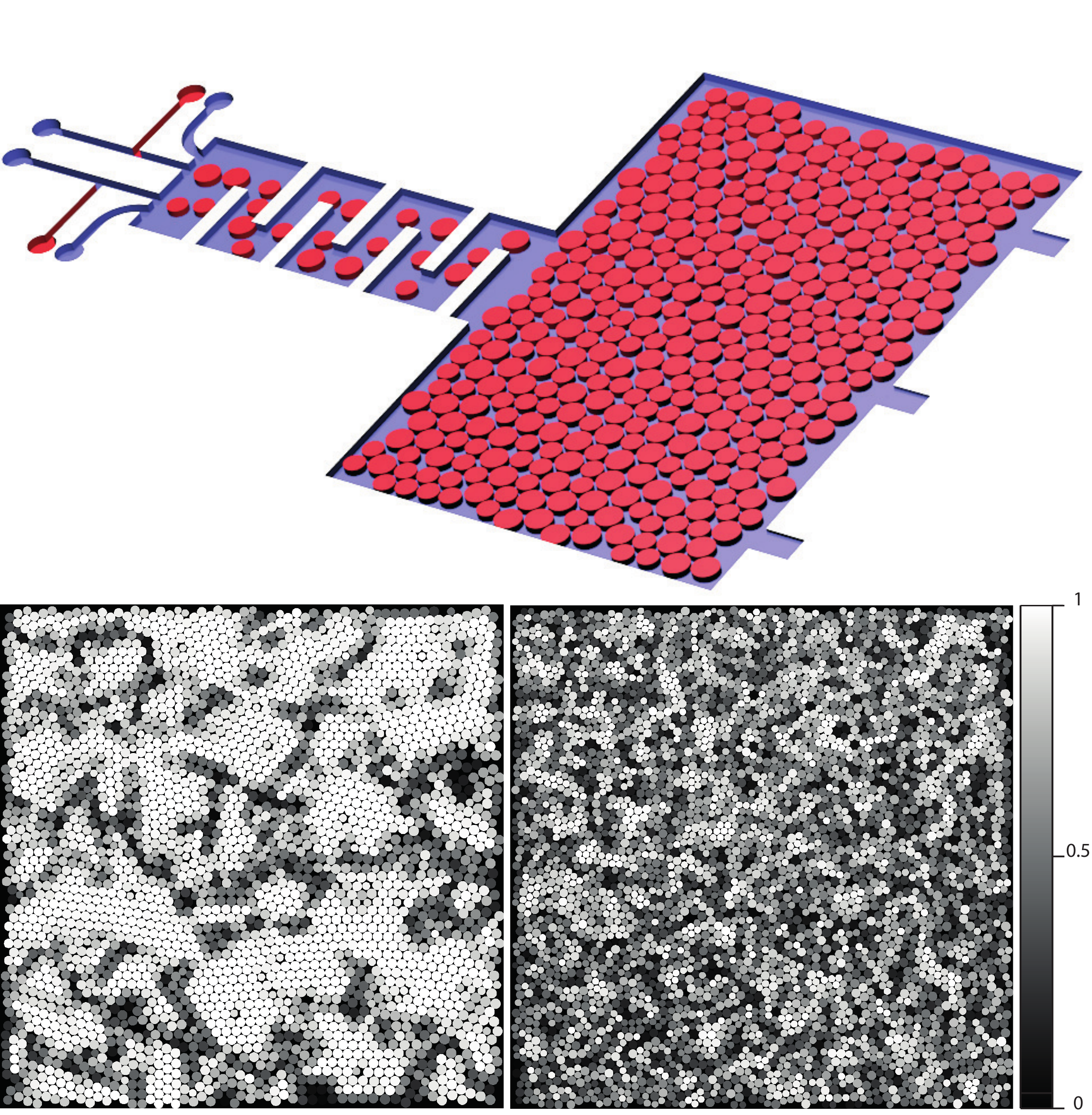}
   \caption{Top. Scheme of the microfluidic chip. Bottom. Experimental images of droplet assemblies.
   Gray scale represents the orientational order parameter $\psi_{6}$. White particles are crystallized in a hexagonal
   lattice. $SR = 0.8$ and $NF = 0.88$ (left, polycristal) or $NF = 0.38$(right, disordered).}
   \label{schemapuceblender}
\end{figure}

Our \new{microfluidic device} shown in Fig.~\ref{schemapuceblender} is made of standard soft photolithography and
replica-molding techniques using polydimethylsiloxane (PDMS). Two populations of  droplets (FC-3283 in 0.2\% SDS
solution) with two different sizes (typically less than 2\% of polydispersity) are created in two separate T-junctions. 
The height of the channels ($\SI{10}{\micro\metre}$) forces the droplet and to adopt a pancake shape. \new{We can
therefore consider the system as 2D disks which organization does not strongly depend on the true thickness of the Hele-Shaw cell \citep{desmond2013experimental}.} The droplet size is in the range
$\SIrange{20}{40}{\micro\metre}$.  \new{The relatively small size of droplets and high enough surface tension ($22 \pm 0.5 \si{\milli\newton\per\metre}$) keep them circular due to Laplace pressure inside while being close to the jamming point.} Droplet production rates, and sizes can be tuned by varying
pressures in four input channels. As a result, the size ratio  between droplets of different sizes
$SR=R_\mathrm{s}/R_\mathrm{b}$ (here $s$ stands for small and $b$ for big droplets) and the number fraction of big
particles $NF=N_\mathrm{b}/N$ are adjustable. Defined this way, both ratios vary from 0 to 1. Then, the droplets pass
through the micromixer which brings disorder and homogeneity to the droplets  by randomizing their entry in the
observation chamber. \new{As entering, they self-organize themselves into a close packing by plastic events. Down from the entrance, plastic events rarefy and the droplets seem to adopt their final configuration with almost
circular shapes~\cite{chen2015experimental}.}  The ceiling of the observation chamber is reinforced with a glass slide
placed inside the PDMS to prevent deformations. Filters are placed downstream to retain droplets inside the cell without stopping fluxes. The filters do not stop the droplets completely but let them
pass slowly, allowing one to renew the packing continuously. Top view microscopy images of $N\simeq 3000$ can be taken \new{allowing to probe wavevectors as small as $q_{min} \sim 1/(D_m \sqrt{N})$, $D_m$ being the mean diameter of disks.}
\new{To avoid any ordering due to the walls, deformation close to the filters at the
exit, and loosely packed zones at the entrance, the photographs are taken at least five layers far from the edges of the
observation chamber~\cite{desmond2009random,chen2015experimental}}. The interval between images is typically on the
order of $30$ seconds, which corresponds to the duration of the filling of the observation chamber. Therefore, systems
are completely \new{refreshed}, and the images are uncorrelated. For each experimental set of parameters at least $30$
images are taken. Statistical measures (e.g. structure factor, fraction of particles with a given coordination) are
calculated on each image and then averaged over the full ensemble. In each experimental set we have checked that the
populations are well mixed and no global gradients of small or big particle concentrations are present.  \new{We also
ensure isotropy of images by checking circular symmetry of their Fourier transformations.}

We also perform numerical simulations of 2D bidisperse jammed assemblies of disks using freely available code based on
the Lubachevsky-Stillinger (LS) algorithm. It is based on a collision-driven packing generation, and is shown to create
highly jammed bidisperse assembly of disks with a low rate of rattlers~\cite{skoge2006packing,atkinson2016static}.

\begin{figure*}
   \centering
   \subfloat{\includegraphics[height=4cm]{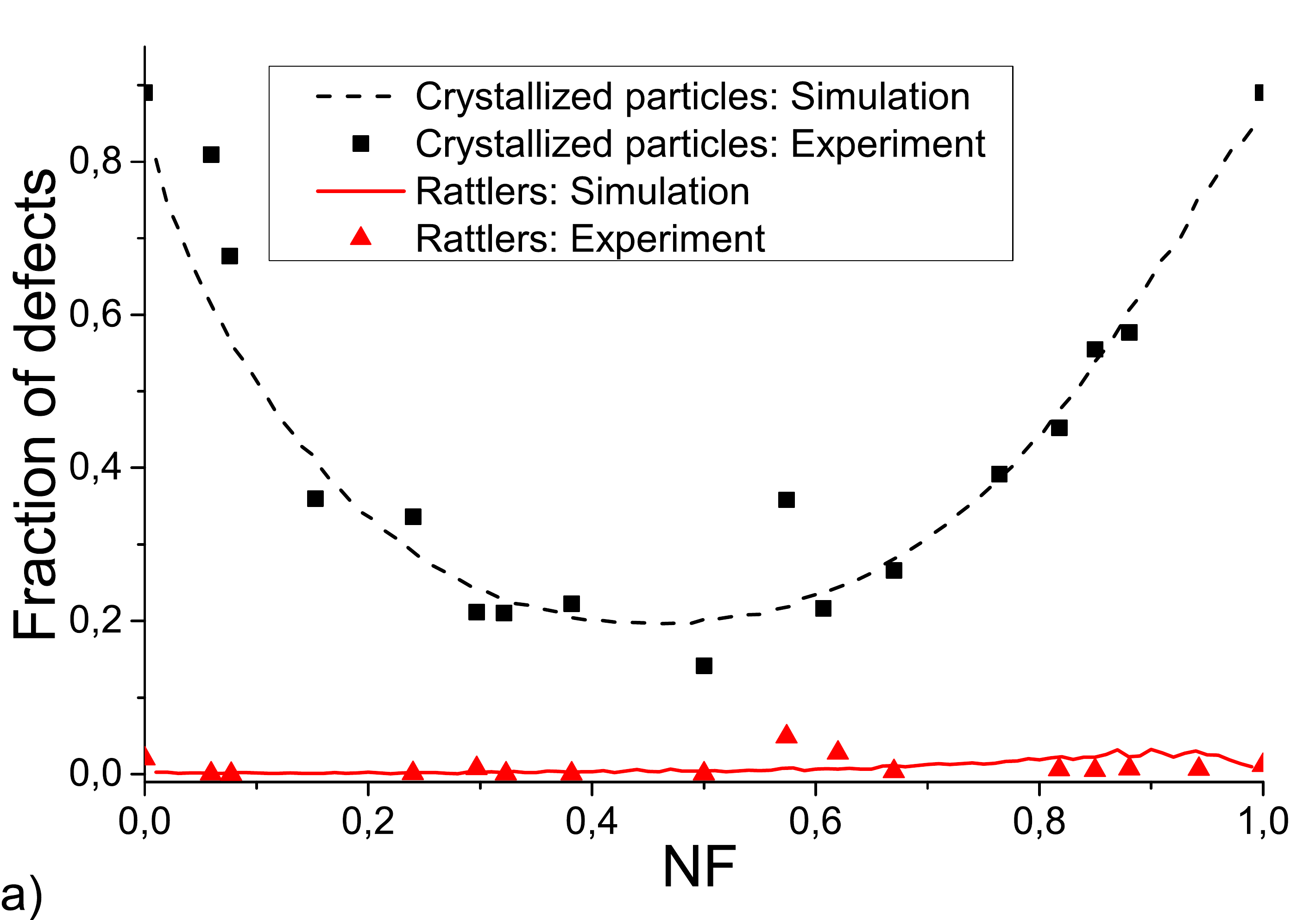}\label{CrystRattlvsnumberfraction}}
   \subfloat{\includegraphics[height=4cm]{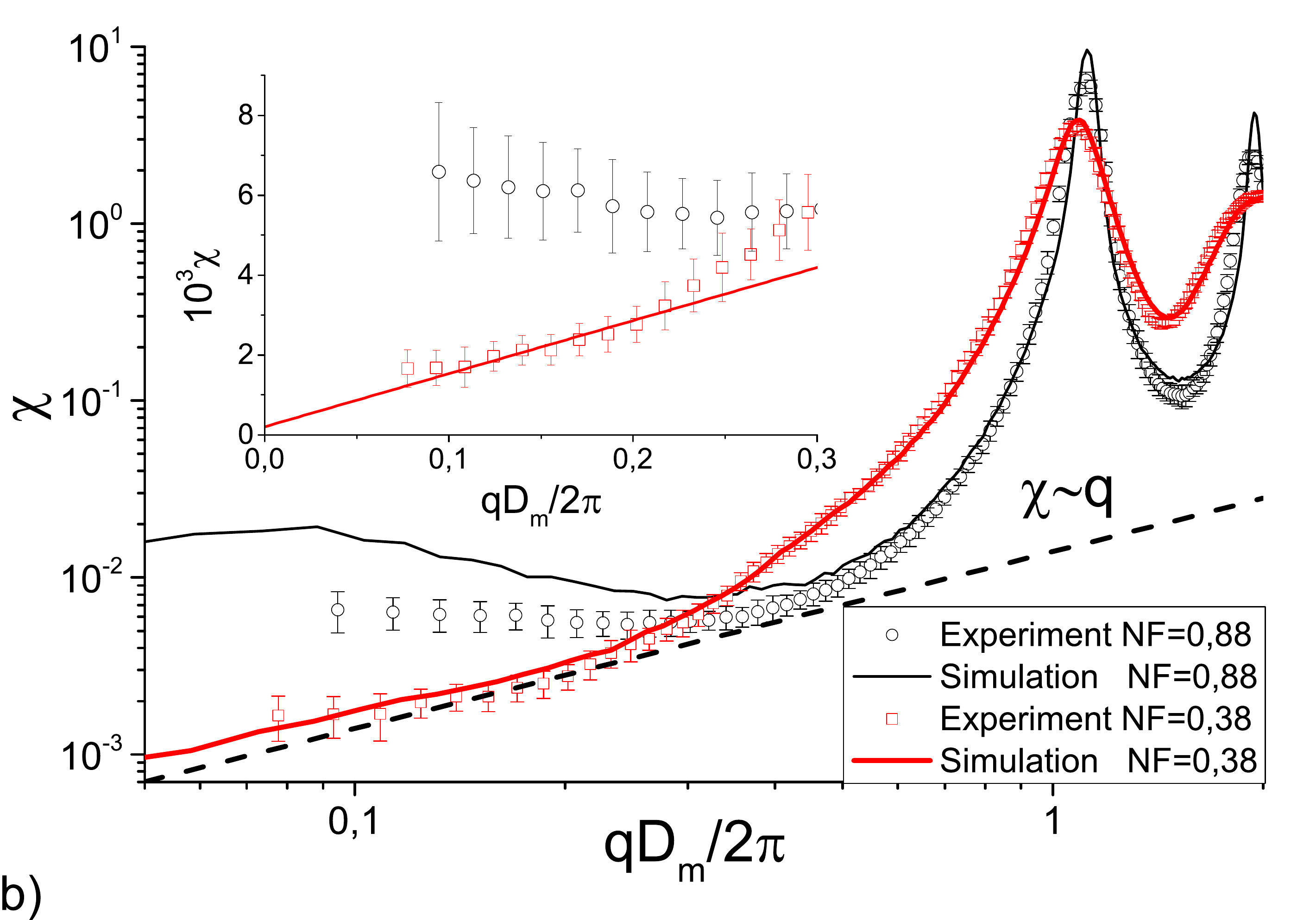}\label{strfactorexpsvssimulation}}
   \subfloat{\includegraphics[height=4cm]{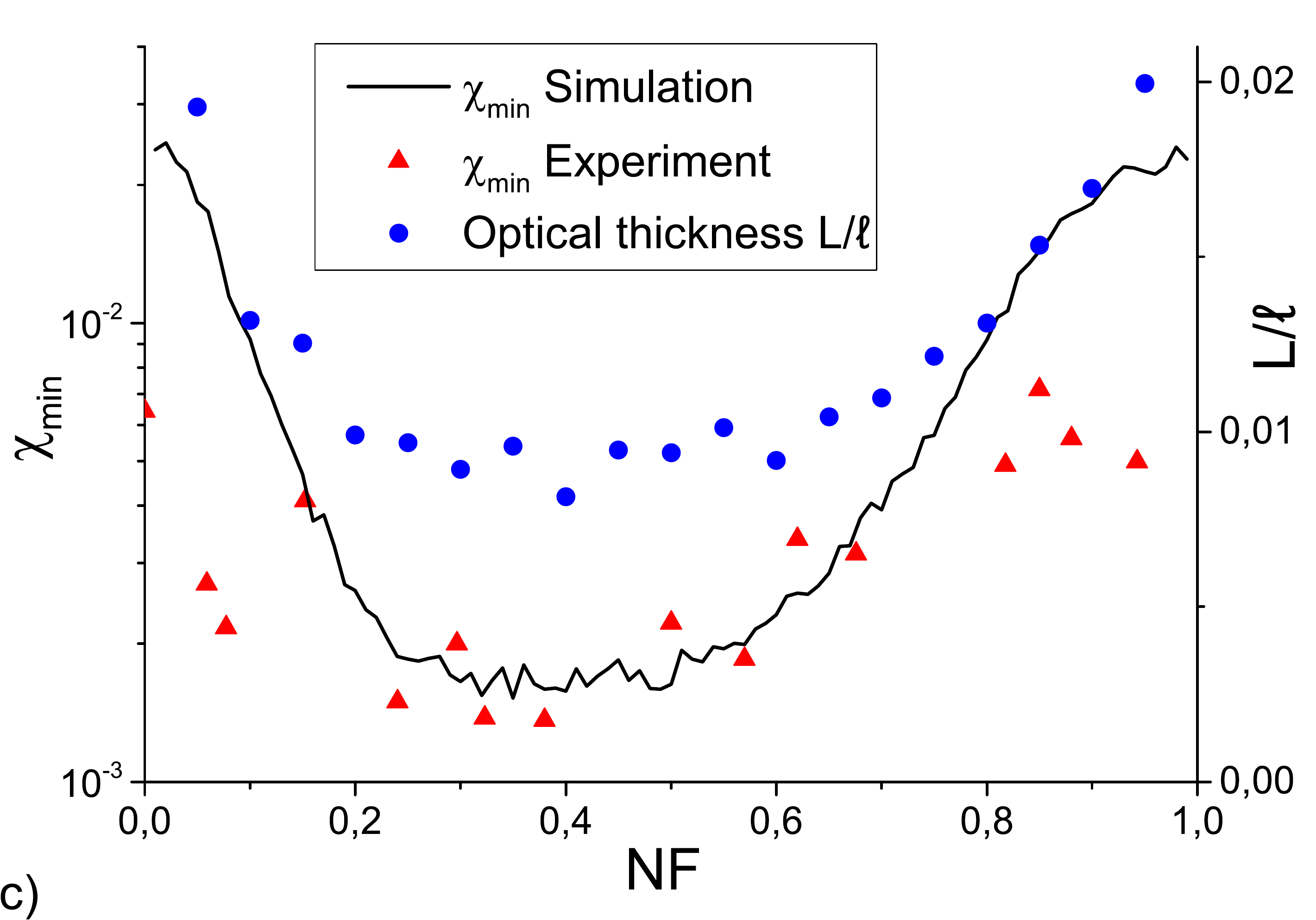}\label{ksiminvsNRSR80}}
   \caption{ \protect\subref{CrystRattlvsnumberfraction} Fraction of crystalline particles  and rattlers for simulated
   and experimental binary mixtures of solid particles for different fraction of big ones $NF$.
   \protect\subref{strfactorexpsvssimulation} Spectral density of a binary disk mixture for different fractions of big
   particles $NF$. Linear dependence $S \sim q$ characteristic of MRJ systems is shown as a guide for eyes. \new{Inset:
   linear fit of experimental data  for $NF=0.38$, y-intercept $\num{2+-3e-4}$.}
   \protect\subref{ksiminvsNRSR80} Spectral density at minimum experimentally available wavevector ($qD_{m}/ \pi \approx
   0.1$) and optical thickness of the simulated 2D jammed assemblies $L/\ell$  as a function of $NF$. The refractive
   index is $1$ for the host medium and $1.3$ for the droplets. A strong correlation between the fraction of crystal
   particles, the minimum spectral density $\chi_{min}$ and optical thickness $L/\ell$ is clearly observed. All figures
   correspond to $SR \approx 0.80$.}
\end{figure*}

To explore the whole diagram of different size ratio and number fraction, we start by keeping the size ratio $SR$
constant, at a value $0.8$ which is a standard value used in most of previous
studies~\cite{donev2006binary,zachary2011hyperuniform}, and varying $NF$.  We observe that for values of $NF$ close to
$0$ or $1$, the system forms a polycrystal with well distinguishable hexagonal crystal domains. For
moderate $NF$ values, we observe disordered structures where crystal domains are small and rare, and the assembly
resembles  a homogeneous mixture of small and big disks. Two experimental pictures for both polycrystal and disordered
systems are shown in Fig.~\ref{schemapuceblender}. To be more quantitative, the experimental pictures are colored using
the bond orientational order parameter $\psi_{6}$, which shows how the weighted Voronoi cell for each particle is close
to a regular hexagon and gives it a score from 0 to 1, approaching 1 for particles with an ideal 6-fold  symmetry. By
considering a particle to be crystallized if $\psi_{6}>0.9$, we plot in Fig.~\ref{CrystRattlvsnumberfraction} the
fraction of crystallized particles both for experimental and simulated samples as a function of $NF$. The data clearly
show that there exists an optimum where the fraction of crystallized particles is minimized. It is close to this point
that MRJ and the most hyperuniform systems are expected. Fig.~\ref{CrystRattlvsnumberfraction} also shows that  the
rate of rattlers, defined as particles having less than three neighbor contacts, is always rather low in our system
(less than 1 \%). Therefore, they should not strongly affect the structural properties. This also indicates that we
indeed succeed in approaching a jamming point. 

To check hyperuniformity of the obtained samples, we calculate the spectral density $\chi (\textbf{q})$, which is an
extension of the classical structure factor for bidisperse and more generally polydisperse particle assemblies. In our
calculations we use the spectral density definition which treats each particle as a point object with weight equal to
the particle area~\cite{wu2015search,berthier2011suppressed}:

\begin{equation}
   \chi (\textbf{q}) = \dfrac{1}{N \left\langle s_{i}^{2}\right\rangle} |\phi (\textbf{q})|^{2},
\end{equation}
where $N$ is the number of particles and $s_{i}$ are their areas. In this expression, $\phi (\textbf{q})$ is the Fourier
transform of the surface fraction defined as
\begin{equation}
   \phi (\textbf{r})= \sum\limits_{i} s_{i}  \delta (\textbf{r}-\textbf{r}_{i}),
\end{equation}
where $\textbf{r}_{i}$ are the positions of  individual particle centers. 

Fig.~\ref{strfactorexpsvssimulation} shows two spectral densities $\chi(\textbf{q})$ in the polycrystal and disordered
cases ($SR \approx 0.8$). The spectral density of a polycrystalline system is not hyperuniform. Indeed,
$\chi(\textbf{q})$ does not monotonously decrease while approaching zero, but rather passes through a local maximum. The
latter can be related to the size and distance between the crystal domains. On the contrary, for disordered systems
close to the minimum of crystallized particle fraction, the spectral density approaches a linear behavior as predicted
for MRJ (see Fig.~\ref{strfactorexpsvssimulation})~\cite{PhysRevLett.95.090604}. \new{A linear fit shown in the inset
reveals a very low spectral density limit $\chi(q=0)=\num{2+-3e-4}$. This confirms hyperuniformity of our
optimized systems \cite{kurita2011incompressibility,berthier2011suppressed}.}

Strictly speaking, hyperuniformity is defined only at infinite lengthscales, which are not relevant for real materials,
nor accessible experimentally~\cite{dreyfus2015diagnosing,kurita2011incompressibility}. To estimate the degree of
hyperuniformity in the finite system, we choose the spectral density $\chi_{min}$ at the minimum wavevector ($qD_{m}/ 2 \pi \approx 0.1$) available from our experimental data. It probes density
fluctuations at large lengthscale about ten particle diameters which are sufficient for the sought optical
properties~\cite{froufe2016role}. \new{Also compared to y-intercept, $\chi_{min}$ does not assume any linear behavior
which is, for instance, disputed in \cite{ikeda2017large}.}  Both experiments and simulations reveal that $\chi_{min}$
has a single minimum at $NF \approx 0.3-0.4$ (Fig.~\ref{ksiminvsNRSR80}) similarly to the previously described
$\psi_{6}$. This result proves the dominating role of crystalline domains in the behavior of the spectral density, and
guides us towards the most hyperuniform systems available experimentally for a given $SR$.

We also use numerical simulations of light scattering to demonstrate that the optimized bidisperse emulsions are good
candidates for the production of high-density and transparent disordered materials.  In order to explore a range of
scattering wavevectors $\bm{q}$ minimizing the structure factor, we have chosen $\lambda=\SI{500}{\micro\meter}$
which corresponds to $k_0D_m/(2\pi)=0.12$, so that $|\bm{q}| \sim k_0$ lies in the region where the spectral density
takes very small values.  Light scattering from the numerically generated  2D jammed assemblies (TE polarization,
electric field perpendicular to the plane)  is simulated by solving Maxwell's equations using the coupled-dipoles
method~\cite{LAX-1952}.  We use the Mie theory to compute the scattering cross sections $\sigma_s$ of a
single disk, and replace  each one by an electric point dipole with an effective polarizability giving the same
scattering cross-section as the real droplet.  From the calculation of the average field inside the medium, we deduce
the real scattering optical thickness $L / \ell$ of the structures (see SM for details on the method). In
Fig.~\ref{ksiminvsNRSR80} we observe that scattering is strongly suppressed for samples exhibiting the highest degree of
hyperuniformity. This analysis supports that the microfluidic fabrication technique described in this Letter is an
important step towards the assembly of hyperuniform materials for photonics.
 
Interestingly, we can also remark that while the concentration of crystallized particles is similar for experimental and
simulated systems (Fig.~\ref{CrystRattlvsnumberfraction}), considering $\chi_{min}$ there exists a small discrepancy for
polycrystalline samples (Fig.~\ref{ksiminvsNRSR80}). The latter is affected not only by the fraction of crystallized
particles but also by the size and shape of crystal domains, which are rather sensitive to the generation processe. In
the meantime, production of hyperuniform systems is much more robust, making them even more attractive for potential
applications. 

\begin{figure}
   \centering
   \subfloat{\includegraphics[height=3.1cm]{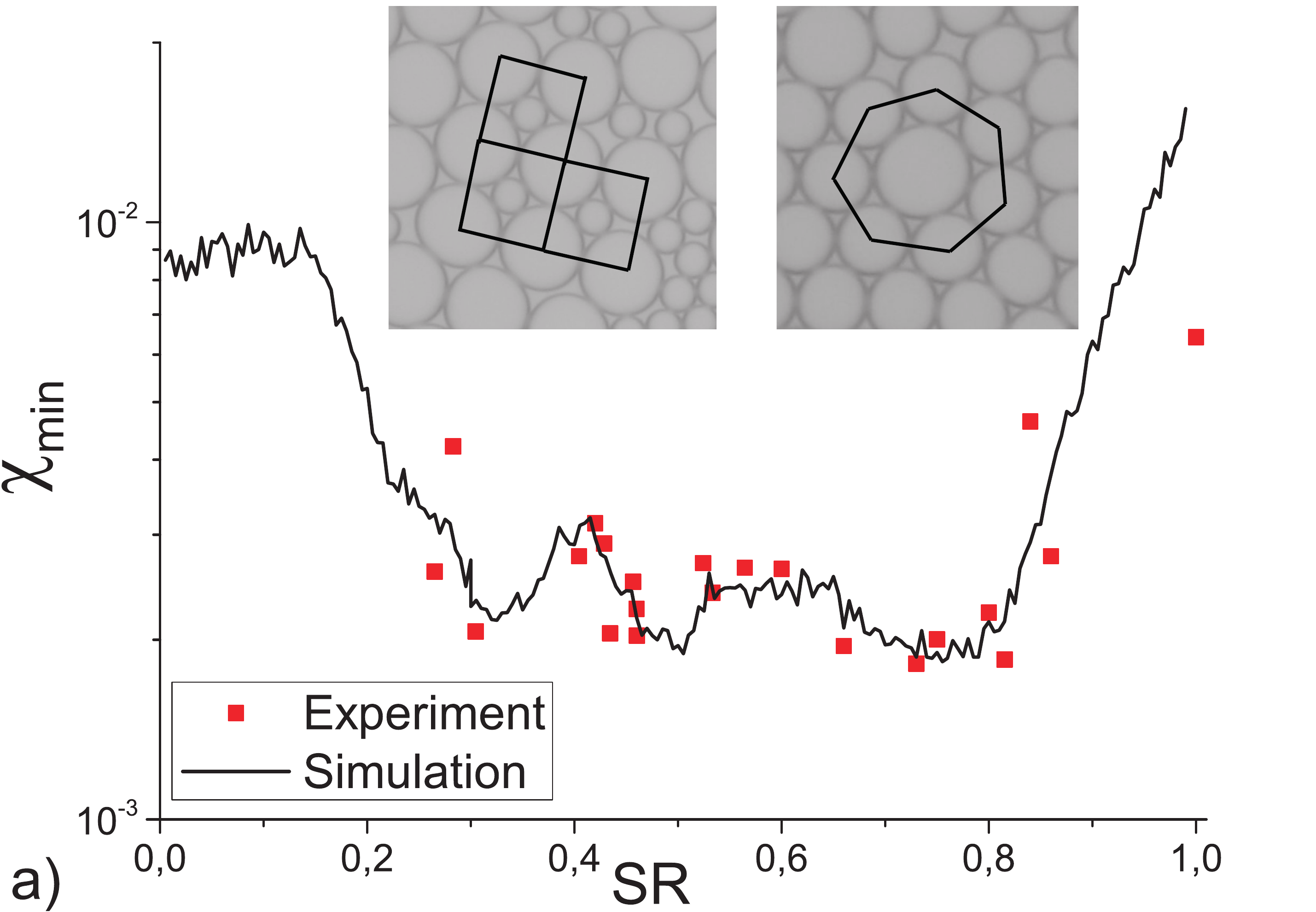}\label{ksiNR50}}
   \subfloat{\includegraphics[height=3.1cm]{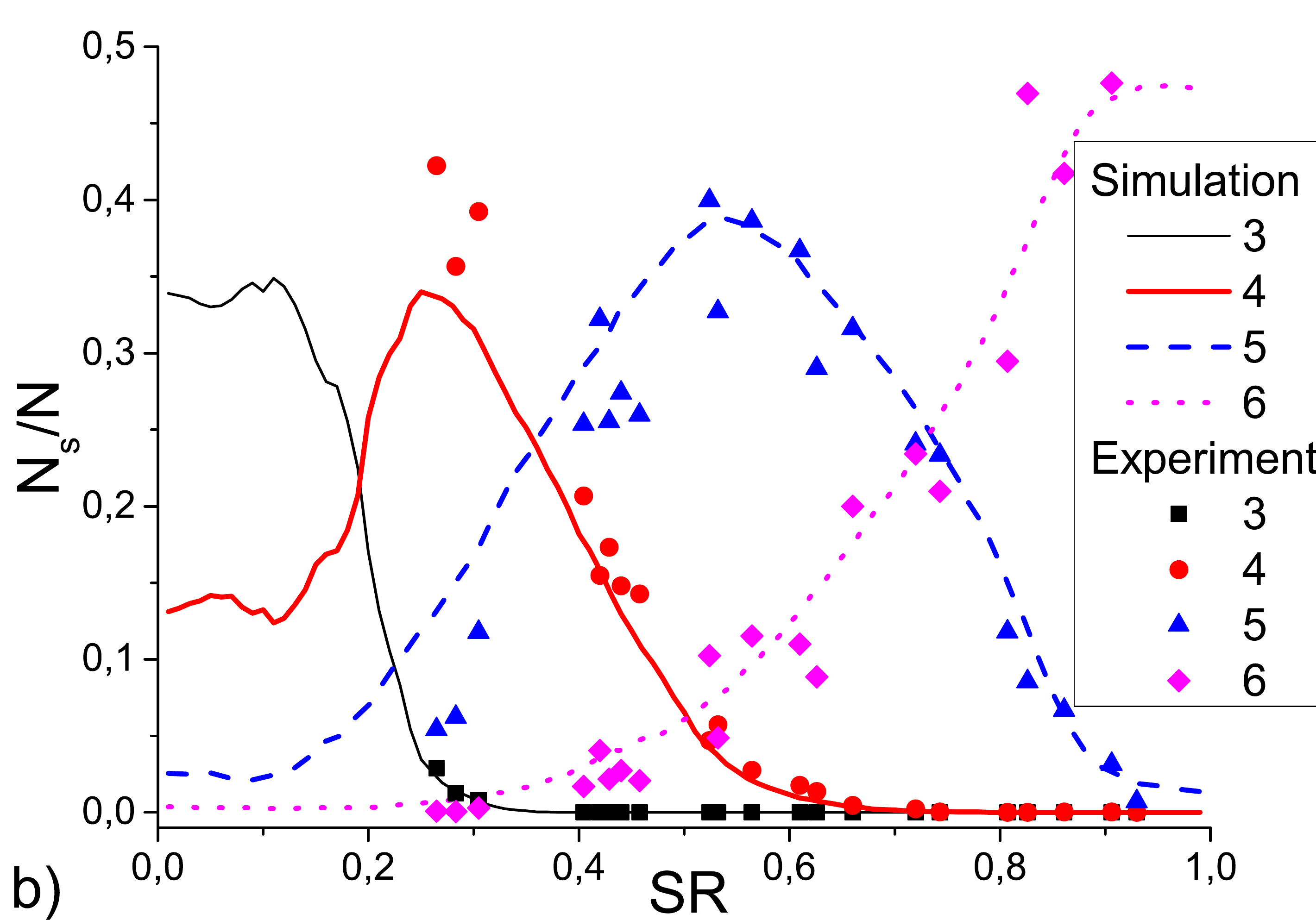}\label{coordinationsmall}}
   \\
   \subfloat{\includegraphics[height=3.1cm]{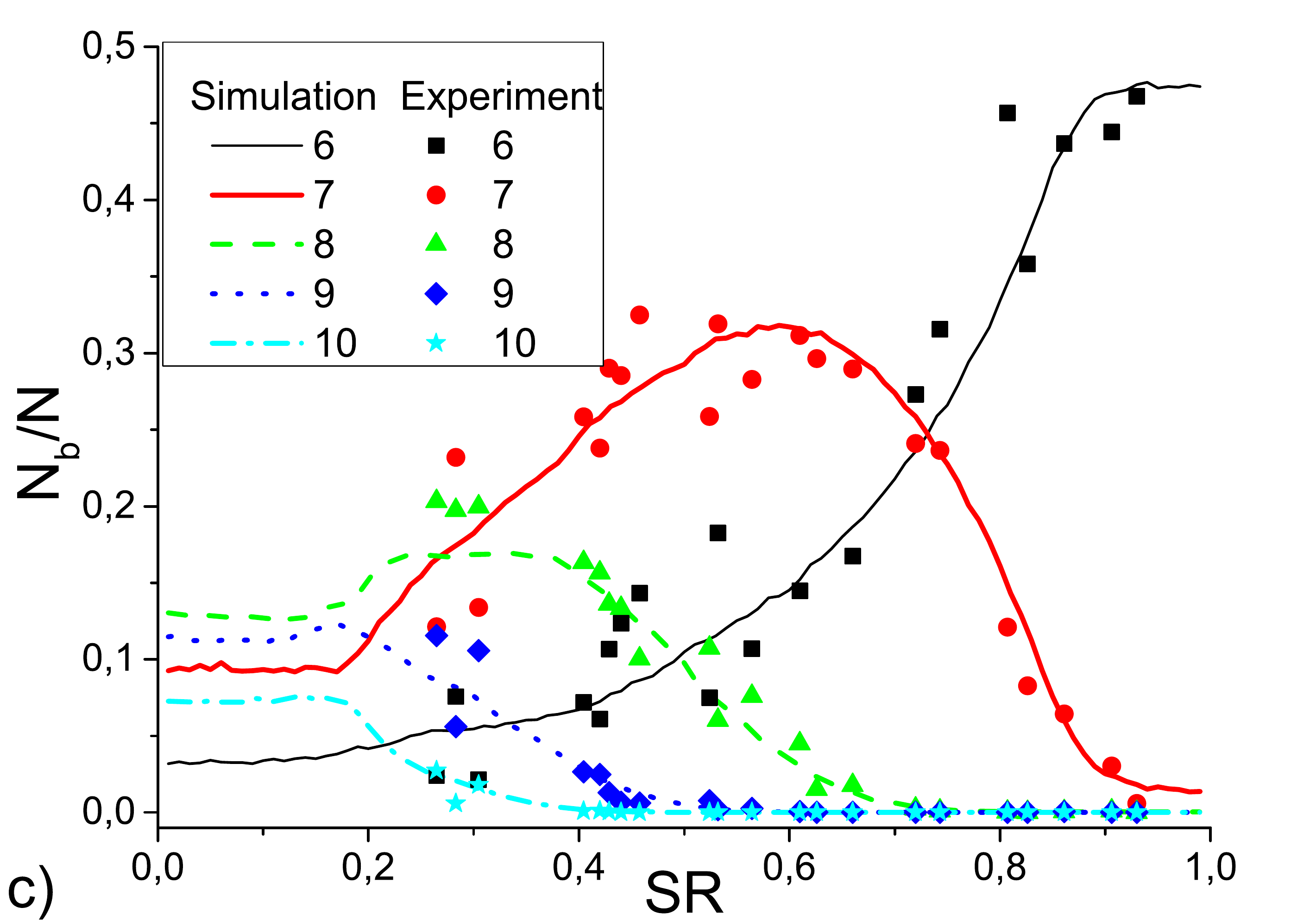}\label{coordinationbig}}
   \subfloat{\includegraphics[height=3.1cm]{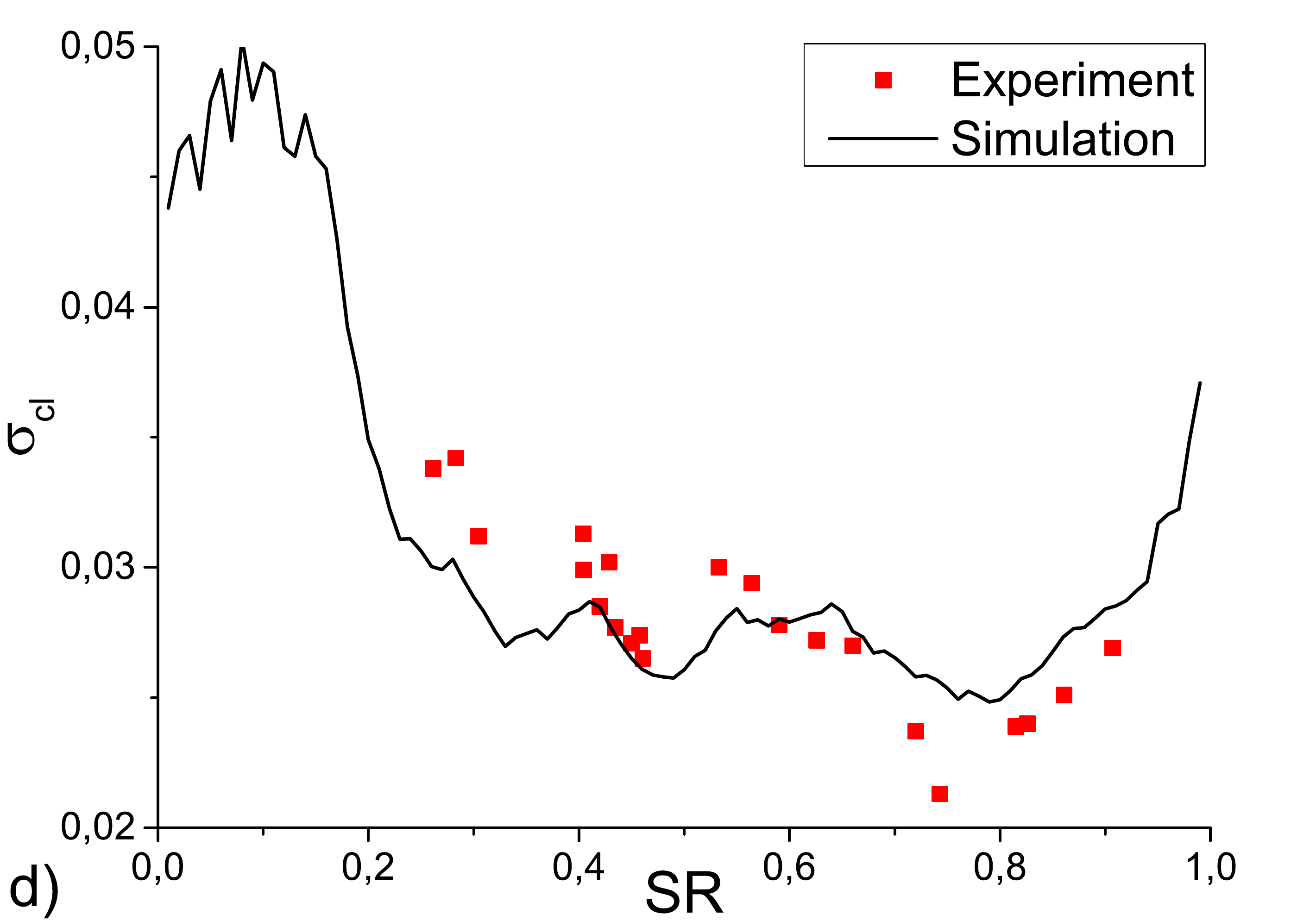}\label{vclf}}
   \caption{ \protect\subref{ksiNR50} Spectral density $\chi_{min}$ at minimum experimentally available wavevector
   ($qD_{m}/ \pi \approx 0.1$). Peaks and valleys are signatures of complex internal organization of different defects.
   Two examples of overpacked defects: a bidisperse crystal domain with four fold symmetry $SR= 0.40$ (left), a cluster
   of a big particle surrounded by seven small particles $SR = 0.70$ (right). \protect\subref{coordinationsmall}
   Fraction of small particles having a given coordination number as a function of size ratio $SR$.
   \protect\subref{coordinationbig} Fraction of big particle having a given coordination number as a function of size
   ratio $SR$.  \protect\subref{vclf} Variance for cluster local fraction. All figures correspond to $NF \approx 0.50$.}
\end{figure}

Surprisingly, we cannot simply extend this correlation between $\chi_{min}$ and $\psi_{6}$ for systems with various size
ratios. Actually, by varying the size ratio for a fixed number fraction $NF=0.5$, we reveal the existence of a few local minima for
$\chi_{min}$, pointing out a much more complex behavior (see Fig.~\ref{ksiNR50}). Several size ratios, ranging from $0.3$ to $0.8$ seem to
have a low rate of density fluctuations. The crystallisation can still account for the general behavior of the system,
namely for the increase of density fluctuations for $SR$ close to $0$ and $1$. But, peaks and valleys for $\chi_{min}$
for medium range of $SR$ (0.3-0.8) cannot be explained solely with the hexagonal crystal domains, and thus
second order defects should be introduced. A similar \new{nonmonotonic} behavior has been previously described for the
surface fraction of simulated bidisperse jammed assemblies, although no explanation has thus far been
provided~\cite{koeze2016mapping,mascioli2017percolation}.

\begin{figure}
   \includegraphics[width=1\linewidth]{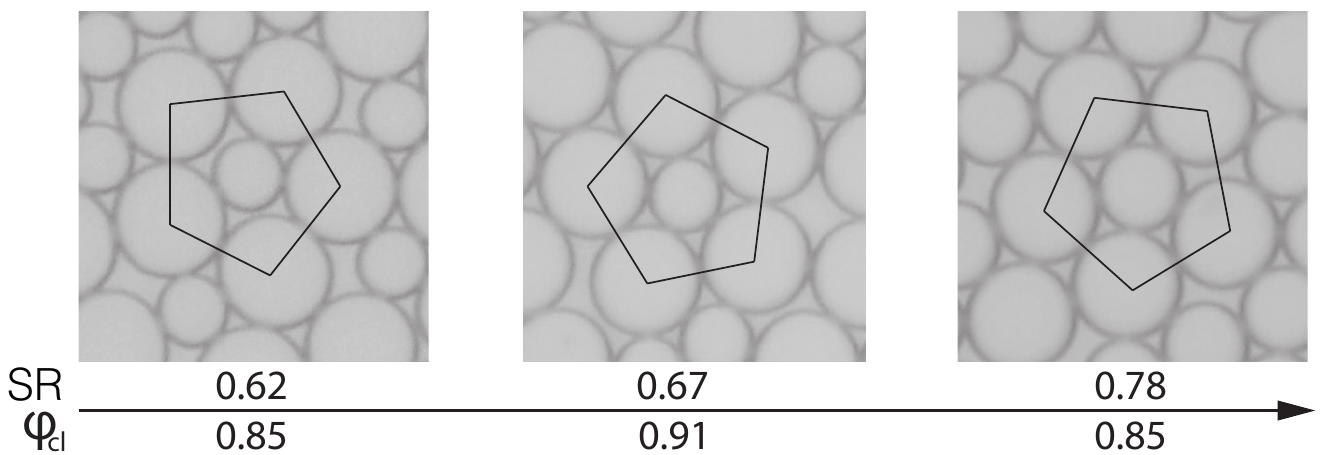}
   \caption{Geometrical resonance for a cluster of one small and five big particles. Three size ratios are shown:
   slightly before, close and slightly after the resonance size ratio $SR \approx 0.7$ corresponding to a compact
   cluster. The compact cluster has a density $\varphi_{cl}$ well above the loosely packed ones.}
   \label{resonance}
\end{figure}

The initial reduction of density fluctuations with decreasing $SR$ from $1$ is explained by the destruction of hexagonal
crystal zones by analogy with $NF$ varying systems. The minimum spectral density $\chi_{min}$ reaches its extremum about
$SR\approx 0.8$. As shown in Fig.~\ref{coordinationsmall} and \ref{coordinationbig}, for size ratio below $0.8$, small
particles with coordination five (number of edges of their weighted Voronoi cell) and big particles with coordination
seven can be identified \new{\cite{mascioli2017percolation}}. The particles under consideration, together with their
closest neighbors, can arrange in compact clusters at appropriate size ratios. Actually, the first type of such
clusters is a big particle surrounded by seven small ones (see Fig.~\ref{ksiNR50}). From simple geometrical
considerations, we can predict the size ratio $SR$ where these clusters appear by calculating the ideal cluster
conformation: $\sin(\pi/7)/[1-\sin(\pi/7)]\approx 0.76$. To give a quantitative estimate of the cluster density, we
define\new{, for each particle,} the cluster fraction $\varphi^{i}_{cl}$ as the ratio of  the occupied surface in the
polygon, whose \new{vertices} are the centers of the surrounding particles, over its area $A_{i}$
(Fig.~\ref{resonance}). For instance, in the considered example, we find a cluster fraction on the order of $0.909$,
slightly higher than that corresponding to hexagonal crystals ($0.906$) and much higher than the average surface
fraction (about $0.84$). Such clusters are geometrical defects that locally increase the density and create additional
fluctuations. 

Each type of cluster is compact only at a particular size ratio $SR$, and exists mainly in the vicinity of this $SR$.
Typically, below and above the characteristic size, the clusters are loosely packed and have a density comparable to the
surrounding medium, while close to the characteristic $SR$ they form overpacked defects (see  Fig.~\ref{resonance}). This
gives a kind of geometric resonance for the spectral density. In the range of $SR$ about 0.5-0.7, a rich morphology of
clusters exists with coordination five and seven for the central particle. For example, we can identify small particles surrounded by two big
and three small particles, or by three big and two small particles, etc. 
In this work, we do not intend to precisely describe the whole variety, although a tentative classification has been
proposed~\cite{kennedy2006compact}. Resonances for different cluster types overlap giving a plateau of $\chi_{min}$. In
comparison to hexagonal crystal domains, these overpacked defects do not have large spatial extensions and are often
limited to one cluster, thus creating only second order fluctuations. 

In Fig.~\ref{ksiNR50}, the peak around $SR \approx0.4$ is particular as it is due to the appearance of four fold symmetry
bidisperse crystal domains. The footprints of this crystal are found in the increased fraction of small particles with
coordination four, and of big particles with coordination eight (see Fig.~\ref{coordinationsmall} and \ref{coordinationbig}). A
peculiarity of these clusters is that they are often organized in small crystallized zones (Fig.~\ref{ksiNR50}). As hexagonal crystal zones,
such domains have higher density and therefore increase the density fluctuations. They are overpacked defects that often
extend to more than one cluster. Actually, in a first approximation (low rate of defects), we can estimate that the
influence of any defect on $\chi_{min}$ scales with its concentration and the surface area square. This power law suggests
that even small crystal zones influence the spectral density much more than the individual clusters. Also other types of
bidisperse crystal domains are expected to appear (see Refs.~\cite{kennedy2006compact,Heppes2003} for some
examples).

If $SR$ decreases further, the four-fold crystals start to be mechanically unstable, and \new{below} $SR=0.15$ they are replaced by
Appolonian packing (small particles in the interstices between three big ones). This packing is clearly observed by the
appearance of a coordination three for small particles and larger than nine for big particles. The big particles are organized in
hexagonal crystal domains which then cause large increase of $\chi_{min}$. Some additional density fluctuations may arise due to the
fact that small particles may occupy only a part of interstices. \new{These small particles in the interstices also show an example of rattlers (particles with less than three neighboring contacts) which can be overpacked defects or even have a cluster fraction close to the mean value. Hence we prefer to describe the system in terms of overpacked and underpacked defects.}

In order to properly monitor the quantity of various clusters, we can calculate for the
set of obtained cluster local fractions $\varphi_{cl}^{i}$ the standard deviation $\sigma_{cl}$ weighted by the cluster area  $A_{i} / \sum A_{i}$. 
The evolution of this quantity versus the size ratio is presented in
Fig.~\ref{vclf} for $NF=0.5$, showing that we are able to recover all peaks and valleys given by $\chi_{min}$. The standard
deviation $\sigma_{cl}$ reports local phenomenon by measuring the disparity of every cluster towards the mean surface
fraction of the system, and, $\chi_{min}$ reports density fluctuations at large length scale. Therefore for jammed
particle assemblies we can correlate hyperuniformity at large lengthscales and the local environment of particles. It
makes $\sigma_{cl}$ also a useful tool to estimate density fluctuations with finite size images. The fact that
hyperuniformity is related to local particle environment also gives expectations to make a statistical description of
the problem using a kind of granocentric model such as in Ref.~\cite{clusel2009granocentric,jorjadze2011attractive}.

\new{To conclude, we have successfully produced large sets of two dimensional bidisperse jammed assemblies with well controlled size and number
ratios.}
 In an optimized range of parameters, structures with reduced large scale density
fluctuations, or equivalently a high level of hyperuniformity, emerge. Thus this work \new{confirms} the relevance of microfluidics for the rapid production of
large scale self-assembling hyperuniform materials \new{\cite{weijs2015emergent}}. The geometrical features of the fabricated
structures are supported by numerical simulations, in very good agreement with the experiments. \new{Our results establish links
between the overall local parameters such as coordination and cluster fraction, and long range density fluctuations.
This can potentially help to suggest hyperuniformity for a much wider range of systems incompatible with long range
analysis.}
\new{The proposed experimental approach also has the potential to be
extended to study 3D packings \cite{clusel2009granocentric,jorjadze2011attractive}.} and could lead to the fabrication of optically dense and
transparent materials, as suggested by light scattering simulations.

\begin{acknowledgements}
   JR and PY contributed equally to this work. We are much grateful to C. Cejas, J. McGraw, P. Chaikin, S. Torquato, O.
   Dauchot and L. Berthier for fruitful discussions and suggestions made along the work. The Microflusa project receives
   funding from the European Union Horizon 2020 research and innovation programme under Grant Agreement No. 664823. This
   work was also supported by LABEX and EQUIPEX IPGG, LABEX WIFI (Laboratory of Excellence within the French Program
   ``Investments for the Future'') under references ANR-10-IDEX-0001-02,ANR-10-LABX-31, ANR-10-LABX-24 and
   ANR-10-IDEX-0001-02 PSL*.
\end{acknowledgements}

%%\bibliographystyle{apsrev}
%\bibliography{biblioarticlehyper2.bib}
%

\end{document}